\documentclass{jltp}

\usepackage{graphicx} 

\title{The Critical State in Type-II Superconductors with Cross-Flow Effects}

\author{A. Bad\'{\i}a and C. L\'opez$^{*}$}

\address{
Dep. de F\'{\i}sica de la Materia Condensada-I.C.M.A., Univ. de Zaragoza, SPAIN
\\$^{*}$Dep. de Matem\'atica Aplicada, Univ. de Zaragoza, SPAIN
}

\runninghead{\bf A. Bad\'{\i}a and C. L\'opez}{\bf The Critical State in Superconductors with Cross-Flow}

\begin{document}

\maketitle

\begin{abstract}
A theoretical framework is presented, which allows to explain many experimental facts related to pinning and cross-flow effects between flux tubes in type-II superconductors. It is shown that critical state principles, in the manner introduced by C. P. Bean for parallel vortex lattices, may be used to describe the observed behavior.
We formulate a least action principle, giving place to a variational interpretation of the critical state. The coarse-grained electrodynamic response of the superconductor is solved by minimizing the magnetic field changes, for a current density vector constrained to belong to some bounded set ($\vec{J}\in\Delta$). It is shown that the selection of $\Delta$ determines the specific critical state model in use. Meaningful choices of $\Delta$ are discussed in view of the related physical mechanisms.


\end{abstract}

\section{INTRODUCTION}

Macroscopic magnetic experiments, which are sensitive to thermodynamic averages over the sample's volume, have been a fundamental tool in the research of type-II superconductors. In particular, one can obtain information on the interaction between flux tubes (vortices) and the underlying pinning structure. Within the framework of Bean's critical state model\cite{bean} (CSM) the experimental data can be processed with remarkable ease. Thus, elementary mathematical manipulations are usually enough for getting important material parameters as the critical current density $J_c$.

Being a phenomenological approach, the CSM is supported by {\em mesoscopic} theories,\cite{richardson} which deal with the complex vortex interactions. The physical basis is that a maximum gradient in the density of vortices is allowed by the underlying pinning structure. This gradient is nothing but the critical value of the coarse-grained current density. When more and more vortices are pushed inwards by the external field pressure, avalanche processes are induced as $J_c$ is exceeded, which relax to new metastable structures with maximum current density. Overdamped relaxation is typically assumed, which relates to the fact that external drive variations occur slowly as compared to the superconductor's diffusion time.

\begin{figure}
\centerline{
\includegraphics[height=2in]{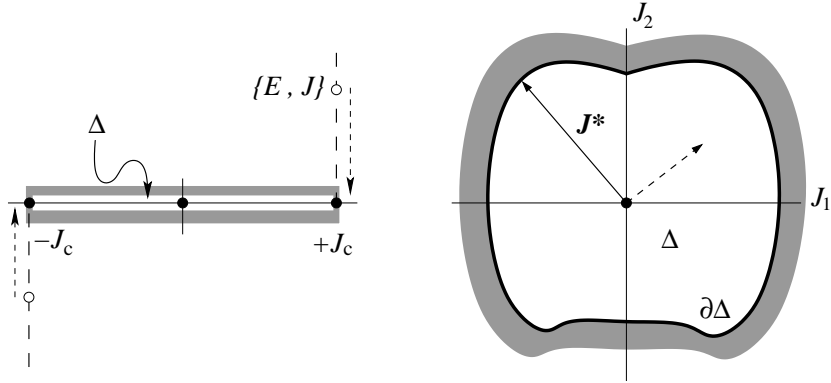}
}
\caption{Restriction sets for the current density ($\vec{J}\in\Delta$)}  
\label{fig:Delta}
\end{figure}

From the macroscopic point of view, this course of events may be modelled by the use of the Maxwell equations and a material law in the form of a vertical $E(J)$ characteristic (see Fig.\ref{fig:Delta}). Any {\em dc} current density may flow within the sample in the absence of electric field for the range $-J_{c}\leq J\leq J_{c}$. Corresponding to a very high level of dissipation for $|J|> J_{c}$, magnetic field changes are introduced by electric field values along the vertical line. Eventually, a new {\em dc} state is reached when $E$ comes to zero again, and we have $J=\pm J_c$ in those regions of the sample affected by changes. 

For one dimensional problems with parallel vortices, the actual behavior of $E$ may be skipped over, and many works on the critical state do not even mention this quantity. In fact, the solution of the CSM consists of choosing among one of the possible states for the current density $\{-J_{c},0,J_{c}\}$ according to the rule that {\em any electromotive force due to external field variations induces the maximum current density flow}. Nevertheless, as soon as multidimensional situations are considered (nonparallel vortices), the problem becomes much more difficult because an infinite number of possible states for the critical current density $\vec{J}_{c}$ are allowed. Several approximations have been introduced over the last decades for dealing with such a situation.

In the early 70's the consideration of simultaneous application of longitudinal magnetic field and transport current along type-II wires led to the concept of flux compression between mutually perpendicular vortices. This approach allowed to explain the so-called {\em paramagnetic effect} related to the increase of magnetic flux towards the center of the sample and basically neglects the flux cutting events between nonparallel vortices.\cite{campbellevetts} 

However, in the following decade, flux cutting was recognized to be the essential mechanism for a reasonable understanding of rotation field experiments\cite{boyer,boyer2,cave} and gave place to the concept of double critical state model\cite{clemgonzalez1} (DCSM). In this approach, the so-called parallel critical current density $J_{c\parallel}$ and perpendicular critical current density $J_{c\perp}$ were introduced. $J_{c\parallel}$ relates to the flux cutting threshold, whereas $J_{c\perp}$ stands for the {\em conventional} depinning current density. 

Recently, a more sophisticated approach was introduced, the so-called {\em two-velocity hydrodynamic model}.\cite{voloshin} Essentially, this theory incorporates the flux pinning and cutting phenomena within the framework of flux line lattices (FLLs) which consist of two vortex subsystems, moving at different speeds. This physical description relates to the mathematical need of a discontinuity in the velocity field in order to allow for crossing events.

Finite element based models are also available,\cite{bosspri} which allow to compute magnetization properties of many non-ideal geometries of crucial importance in the field of applied superconductivity. These are typically in the form of variational statements and, to our knowledge have been exploited for scalar $E(J)$ relations (i.e.: $\vec{E}\parallel\vec{J}$ is assumed).

A variational approximation to the critical state was also suggested in Ref.\onlinecite{bhagwat}, by a guessed {\em minimum flux change} criterion. The authors of that work argued that, in certain conditions ($\vec{E}\parallel\vec{J}$), successive field penetration profiles in the magnetization process could be obtained by minimizing the squared flux density variation for constant current density modulus.

In the theory presented in this paper we propose a generalization, which incorporates a vector {\em vertical} $\vec{E}(\vec{J})$ law, but in the manner of Bean's approach still avoids the explicit use of induced electric fields. A principle is provided for determining the critical current density vector $\vec{J}_c$ within the sample, subsequent to the breaking of equilibrium by the external drive. We keep the main physical ingredient that infinite dissipation occurs if some critical value is exceeded. As this value is a vector quantity a new mathematical framework is required, so as to find the actual critical state in which the system settles subsequent to a specific manipulation.

\section{MACROSCOPIC FIELD EQUATIONS}

In this section we will define the macroscopic fields involved in the CSM and their relation to measurable quantities. Our work has been developed within the assumption that the sample's response is dominated by the {\em irreversible} contribution. This means that the equilibrium response of the FLL in the absence of pinning centers is neglected. For global magnetization measurements in which the integrated magnetic moment of the sample is measured, this is a very good approximation, unless for weak pinning specimens (see Ref.\onlinecite{krasnov} and references therein). Then, the coarse-grained electrodynamics if formulated in terms of
\begin{itemize}
\item[(i)] The {\em flux density} $\vec{B}(\vec{x})$ within the sample is the average of the microscopic field intensity $\vec{h}$.
\item[(ii)] The {\em magnetic field} $\vec{H}(\vec{x})$, in the absence of equilibrium magnetization, is linearly connected to $\vec{B}$ by $\vec{B}=\mu_{0}\vec{H}$.
\item[(iii)] The averaged current density may be calculated from $\vec{J}=\nabla\times\vec{H}$.
\item[(iv)] The electric field can be evaluated by $
\nabla\times\vec{E}=-\mu_{0}\partial_{t}\vec{H}$.
\end{itemize}
Finally, the connection to observable quantities is done by characterizing the external field sources and the sample's response.
\begin{itemize}
\item[(v)] On neglecting finite size effects, the magnetic source ($\vec{H}_{S}$) enters as a boundary condition for the flux density at the surface of the sample. The tangential component is continuous $B_{T}(surface)=\mu_{0}H_{S,T}$.
\item[(vi)] Then, the measured magnetic moment of the sample per unit volume is $\vec{M}\equiv\langle\vec{B}\rangle /\mu_{0}-\vec{H}_{S}=\langle\vec{H}\rangle -\vec{H}_{S}$.
\end{itemize}
For further use we notice that, in the case of one independent variable, Ampere's law allows the following representation
\begin{equation}
\label{eq:ampere}
\vec{J}=\nabla\times\vec{H}=\frac{dH}{dx}\hat{\alpha}-H\frac{d\alpha}{dx}\hat{H}
=J_{\perp}\hat{\alpha}+J_{\parallel}\hat{H}
\end{equation}
where $\alpha$ is the angle between the magnetic field and the $Z$ axis, and $\hat{H}$ and $\hat{\alpha}$ are unit vectors respectively directed along $\vec{H}$ or perpendicular.

\section{OPTIMAL CONTROL THEORY FOR THE CRITICAL STATE}

In this section we show that Bean's ansatz for parallel FLL's allows a {\em geometrical} interpretation, which one can straightforwardly generalize for complex systems with nonparallel vortices. Recall that penetrating flux profiles in the slab geometry may be obtained by solving the equation
\[
\frac{dH}{dx}=\pm J_{c}\; , \; 0
\]
where the sign selection is made according to the rule: {\em any electromotive force due to external field variations induces the maximum current density flow}.

In simple terms, one is just using Lenz's law. Thus, the macroscopic electrodynamics of a hard type-II superconductor may be viewed as follows. Owing to physical limitations imposed by the material properties, the current density could take any value within the interval $\Delta = [-J_{c},J_{c}]$, but as electromagnetic induction operates all along the process, one gets $J=\pm J_{c}$, according to the most effective way of shielding field variations. From our point of view, the physical limitations within the superconductor define a region in $J$-space where the current density could settle ($J\in\Delta$). Lenz's law establishes that only the boundaries will be occupied (or the point $J=0$ for regions where electric fields have not reached).

For higher dimensional systems, i.e.: vortices are no longer straight parallel lines or $\vec{J}$ is truly a vector [see Eq.(\ref{eq:ampere})], the allowed region for the current density will be a subset of $\vec{J}$-space. In this case, Lenz's law is insufficient for determining the points where the current density lies along the magnetization process. Faraday's law in full form must be appealed. In brief, our critical state theory includes (i) a material law of the kind $\vec{J}\in\Delta$ and (ii) a variational statement of Faraday's law. Below, it will be shown that calling on (ii), the solution of the problem $\vec{J}^{\, *}$ only takes values on the boundary of $\Delta$ (or 0). This has been sketched in Fig.\ref{fig:Delta}. Thus, the {\em essence of the critical state} in superconductors, is defined by the condition
\begin{equation}
\label{eq:boundary}
\vec{J}^{\,*}\in\partial\Delta\qquad ({\rm or}\quad 0) \; .
\end{equation}
The determination of the specific values $\vec{J}^{\, *}(x)$ for any process is described in the following subsection. A brief discussion on the physical basis of the mathematical statement is also given.

\begin{figure}
\centerline{
\includegraphics[height=1.75in]{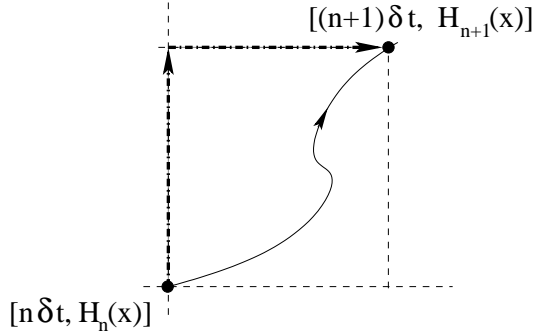}
}
\caption{Quasistationary evolution approximation}  
\label{fig:quasista}
\end{figure}

\subsection{General statement}

As it was shown in previous work\cite{badia1,badia2}, the evolutionary magnetic processes in hard type-II superconductors may be described by quite general principles of linear irreversible thermodynamics. Let us consider a superconductor in some equilibrium state within the experimental range of interest. This means that the system of vortices is at rest by compensation of mutual interactions, pinning forces and the external field pressure. From the macroscopic point of view, the equilibrium states will be characterized by a magnetic field or current density pattern within the sample. For instance, one can define it by a vector function $J_{c_i}(x)$. Consider a small perturbation that drives the system out of equilibrium. Then, a non-equilibrium flow and linearly related internal forces will arise. Specifically, the critical current density has been exceeded by some amount $J_{i}-J_{c_i}$, linearly related to associated electric field components $E_j$. Stationary states in which the external drive keeps time independent values of $\vec{E}$ can be straightforwardly solved by the principle of minimum global entropy production, introduced by Prigogine in the context of linear irreversible thermodynamics\cite{prigogine}. In this case, one can write it as
\begin{equation}
\label{eq:prigo}
\int_{\Omega}\! \vec{E}\cdot (\vec{J}-\vec{J}_{c}) \qquad \to \quad {\rm min}  \; ,
\end{equation}
where $\Omega$ denotes the volume of the sample. The reader can check that in the particular case $\vec{J}_{c}=0$, this principle leads to the well known current distribution for the stationary regime in normal metals.\cite{landau}

However, we are interested in the time evolution between equilibrium states induced by successive magnetic field changes. One can achieve an appropriate description just by adding up a term to Eq.(\ref{eq:prigo}) which accounts for the magnetic field inertia, in the so-called {\em quasistationary} approximation. The physical idea is schematically shown in Fig.\ref{fig:quasista}. We assume a time-discretized picture in which $n\delta t$ denotes the $n$-th time layer after $n$ equal time steps with increment $\delta t$. Then, for small enough values of $\delta t$, one can also suppose that changes in the magnetic field configuration $\vec{H}_{\rm n}(x)$ are small and substitute the continuum trajectory of the system by a polygonal approximation in which (i) $\vec{H}_{\rm n}(x)$ changes at fixed time to $\vec{H}_{\rm n+1}(x)$ ({\em inertia} term) and (ii) time is incremented for $E_{i}=\rho_{ij}(J_{j}-J_{c_j})=\rho_{ij}[(\nabla\times\vec{H}_{\rm n+1})_{j}-J_{c_j}]$ ({\em stationary} entropy production term). Then, the configuration $\vec{H}_{\rm n+1}(x)$ may be obtained from the previous one by the minimization rule
\begin{equation}
\label{eq:fullminim}
\mu_{0}\int_{\Omega}\! |\vec{H}_{\rm n+1}-\vec{H}_{\rm n}|^{2}
+\delta t\int_{\Omega}\! \vec{E}\cdot(\vec{J}-\vec{J}_{c})
\qquad \to \quad {\rm min}  \; .
\end{equation}
The rigorous proof that the previous principle is equivalent to Faraday's law in discretized form plus a constitutive relation of the kind $E_{i}=\rho_{ij}(J_{j}-J_{c_j})$ is straightforward by following the calculations in Ref.\onlinecite{badia2}.

The final form of our principle, which connects to the geometrical interpretation described in the previous paragraph is attained by considering vertical $\vec{E}(\vec{J})$ laws, i.e.: infinite flux flow resistivity. Then, in order to avoid adding a divergent positive quantity, the entropy production term is replaced by the constraint $\vec{J}\in\Delta$. We end up with the principle

\vspace{0.25cm}

{{\hspace{-.5cm}\fboxsep
1.ex\fbox{\parbox{0.95\textwidth}{ 
{\em In a type-II superconducting sample $\Omega$ with an initial
field profile $\vec{H}_{\rm n}(\vec{x})$, and under a small change of 
the external drive, the new profile $\vec{H}_{\rm n+1}(\vec{x})$ minimizes the 
functional}
\begin{equation}
\label{eq:varpri}
{\cal C}[\vec{H}_{\rm n+1}(\vec{x})]
=\frac{1}{2}\int_{\Omega}\! | \vec{H}_{\rm n+1} -
\vec{H}_{\rm n} |^{2} \; ,
\end{equation}
{\em with the boundary conditions imposed by the external source, and 
the constraint $\nabla\times\vec{H}_{\rm n+1}\in\Delta$.}
}}}}

\vspace{0.25cm}

In general, $\Delta$ is a bounded region where the {\em dc} current density vector should remain. As it will be shown in the next subsection, the selection of this region corresponds to using a given critical state model. 

The kind of mathematical statement posed in Eq.(\ref{eq:varpri}) may be solved by several methods. The authors\cite{badia1,badia2} have preferred to use the so-called {\em Optimal Control} (OC) theory,\cite{oc} a powerful generalization of variational calculus, which allows to deal with a wide class of constrained problems. For the readers' benefit we have included a brief survey of this mathematical tool as an appendix of this article. A reminder of the application to the critical state problem is given below.

For the slab geometry, with applied magnetic field parallel to the faces, the current density restriction may be written
\begin{equation}
\label{eq:controlsystem}
{{d{\vec H}_{n+1}}\over {dx}} = {\vec f}({\vec H}_{n+1}, {\vec u},x)\; , \quad
{\vec f} \in \Delta_{\perp} \; .
\end{equation}
This is called control system. Hereafter, we take the X axis perpendicular to the slab faces and the origin of coordinates at the midplane. By construction, the  vector ${\vec f}=(0, f_y,f_z)$ is orthogonal to the physical variable $\vec J$, $f_y = J_z$, $f_z = - J_y$. Notice that, owing to the applied rotation, we use the notation $\Delta_{\perp}$ for the allowed control set. Note also that the function $\vec f$ displays dependence on the local magnetic field (this 
allows to include the usual models $J_c(|{\vec H}|)$,\cite{badia1,badia2} and possible anisotropy), on the position (for potentially  inhomogeneous materials) and on some independent coordinates $\vec u$, the control variables in OC terminology, parametrizing the region $\Delta_{\perp}$.

Next, we define a Hamiltonian density, containing the Lagrangian density to be minimized
\begin{equation}
\label{eq:hamiltonian}
{\cal H}( {\vec H}_{n+1}, {\vec u}, {\vec p}, x)\equiv{\vec p}\cdot {\vec
f}-{{1}\over {2}}| {\vec H}_{n+1} - {\vec H}_n|^2 \; .
\end{equation}
Denoting by ${\vec H}_{n+1}^*(x)$, ${\vec p}^{\,*}(x)$ and ${\vec u}^{\,*}(x)$ the optimal solution functions (i.e. minimizing ${\cal C}$ and satisfying the control system), the OC equations are
\begin{equation}
\label{eqnham1}
{{d{\vec H}_{n+1}^*}\over {dx}} =  {{ \partial{\cal H}}\over
{\partial {\vec p}}} = {\vec f}
({\vec H}_{n+1}^*, {\vec u}^{\,*},x) \; ,
\end{equation}
the adjoint equations for the momenta
\begin{equation}
\label{eqnham2}
{{d{\vec p}^{\,*}}\over {dx}} = - {{\partial {\cal H}}\over {\partial
{\vec H}_{n+1}}}=
{\vec H}_{n+1}^*-{\vec H}_n(x)
   - {\vec p}^{\,*}\cdot {{\partial {\vec f}}\over {\partial {\vec H}_{n+1}}}
({\vec H}_{n+1}^*, {\vec u}^{\,*}, x) \; ,
\end{equation}
and the algebraic condition of maximality
\begin{equation}
\label{eqnmax}
{\cal H} ( {\vec H}, {\vec u}^{\,*}, {\vec p}, x) \geq {\cal H} ( {\vec
H}, {\vec u}, {\vec p}, x)
\qquad \forall \;{\vec f}({\vec H}, {\vec u},x) \in \Delta_{\perp} \; .
\end{equation}

\begin{figure}
\centerline{
\includegraphics[height=2in]{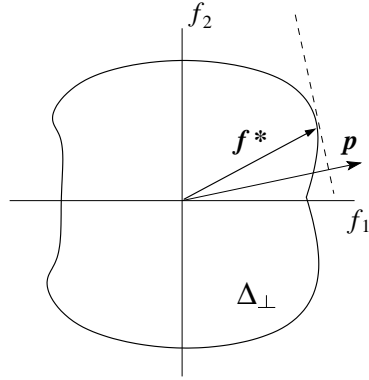}
}
\caption{Maximum projection rule ($\Rightarrow \vec{J}^{\, *}\in\partial\Delta$)}  
\label{fig:projection}
\end{figure}

For the class of Hamiltonian $\cal H$ described above, the algebraic 
condition of maximality is fulfilled for a vector $\vec f$ with 
maximum projection over the momentum $\vec p$. As it apparent in Fig.\ref{fig:projection} and in the various examples of the following subsection, this rule is nothing but the result announced in Eq.(\ref{eq:boundary}): $\vec{J}^{\,*}\in\partial\Delta$. The actual solution $\vec{J}^{\; *}(x)$ can be determined by the Hamiltonian equations displayed above. 

As regards the boundary conditions that must be used for obtaining 
the solution profiles ${\vec H}_{n+1}^*(x)$ from Eqs.(\ref{eqnham1}) 
and (\ref{eqnham2}), several considerations must be made. Firstly, in 
the absence of demagnetizing effects, ${\vec H}_{n+1}^*$ is 
determined on the faces of the slab by continuity of the external 
applied field. For the remaining boundary conditions, two typical 
situations appear. In the first case, the modified penetrating 
profile ${\vec H}_{n+1}^*(x)$ equals the former profile ${\vec H}_n(x)$ 
before reaching the centre of the slab. This gives place to an 
(unknown in advance) point $x^*$ such that ${\vec H}_{n+1}^*(x) = 
{\vec H}_n(x) \quad \forall\; x \leq x^*$. The algebraic free boundary 
condition ${\cal H}(x^*)=0$ applies in this case (see \ref{a1}), and allows to determine $x^*$. In the second case, the new profile never meets the 
former one, and the value of ${\vec H}_{n+1}^*(0)$ is unknown; then, 
the corresponding transversality condition ${\vec p}(0) = {\bf 0}$ (see also \ref{a1}) completes then the number of required boundary conditions.

\subsection{Particular models}

\begin{figure}
\centerline{
\includegraphics[height=3.55in]{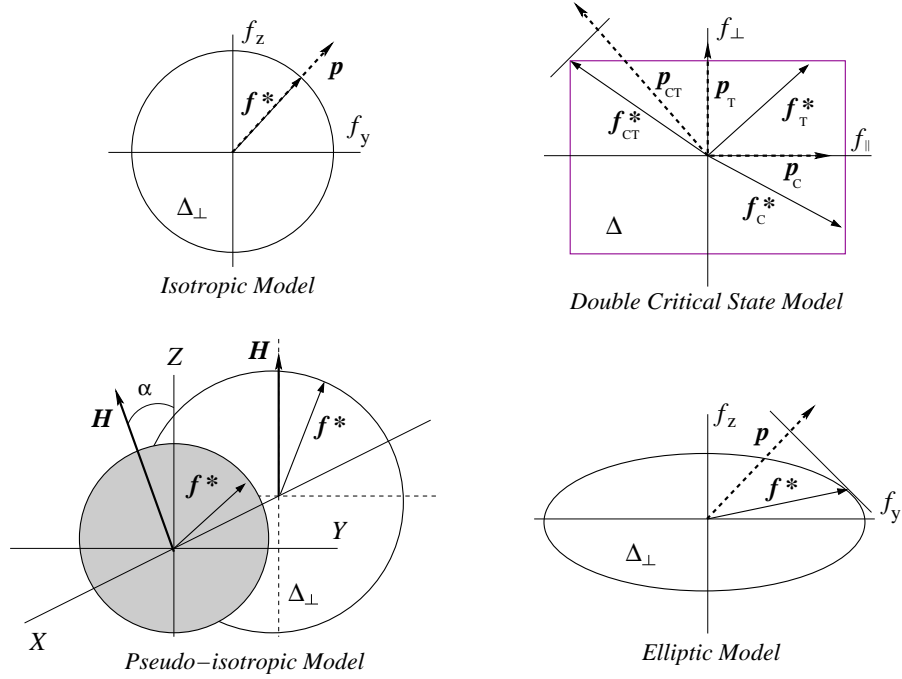}
}
\caption{Critical State Models arising from the selection of the region $\Delta$}  
\label{fig:regions}
\end{figure}

Next, we analyze the models which arise for several selections of the restriction region in Eq.(\ref{eq:controlsystem}). Fig.\ref{fig:regions} depicts such regions and the relation of the functions $\vec{f}$ with the associated momenta in the theory. In this section we will focus on the formal aspects of the solutions. Later on, the relation to the experimental facts will be established. 

\subsubsection{Isotropic model}

The simplest restriction set which one can imagine for two dimensional systems corresponds to choosing $\Delta$ as a circle. Following the notation introduced in Eq.(\ref{eq:controlsystem}), the control system takes the form
\begin{equation}
\label{eq:controlisotropic}
{{d{\vec H}_{n+1}}\over {dx}}={{\vec f}({\vec H}_{n+1},{\vec u},x)=\vec u}\; .
\end{equation}
In this case, ${\vec u}$ is a vector within the unit disk D. For simplicity, dimensionless units have been used: $x$ is given in units of the slab half-thickness $a$, ${\vec J}$ in units of $J_{c}$, and the magnetic field ${\vec H}$ in units of $J_{c}a$. Also for simplicity, here and in what follows we take $J_{c}$ as a  constant.

The maximality condition (maximum projection of $\vec f$ over $\vec p\;$) leads to the solution $\vec{f}^{*}=\vec{u}^{\,*}=\vec{p}^{\,*}/p^{*}$. Note that this means that $\vec{f}^{*}$ is a unit vector always parallel to the momentum as indicated in Fig.\ref{fig:regions}. Eventually, the penetration profiles $H_{\rm n+1,i}^{*}(x)$ can be obtained from the system
\begin{eqnarray}
\frac{dH_{\rm 
n+1,i}^{*}}{dx}&=&\frac{p_{i}^{*}}{p^{*}}
\label{eqncaniso:a}
\\[1ex]
\frac{dp_i^{*}}{dx}&=&H_{\rm n+1,i}^{*}-H_{\rm n,i} 
\label{eqncaniso:b}
\end{eqnarray}
on supplying the appropriate boundary condictions.

The most relevant physical property related to this model is that the maximum allowed current density modulus $J_c$ is carried within those regions which have been affected by the perturbation ($\vec{J}^{\; *}\in\partial\Delta$ in our geometrical language). Then, when a considerable amount of current is required along a given direction (f.i.: $J_z$) for shielding a specific field variation ($H_y$ in this case), the other component of $\vec{J}$ is depressed ($J_y$ in this case). This reduces the shielding capability in the orthogonal field direction ($H_z$ in this case). As a consequence, one can effectively increase the penetration of a given magnetic field component, just by manipulating the perpendicular one. This phenomenon has been named {\em magnetization collapse}\cite{baltaga} and has been widely illustrated in the framework of our theory in Ref.\onlinecite{badia2}.

\subsubsection{Double critical state model}

Below, we show that the DCSM equations can be readily obtained within the framework of our theory. A specific selection of the restriction region $\Delta$, together with the variational principle, render the DCSM as a particular case of our formulation. 

Recall that the flux depinning and cutting thresholds are defined by the critical state conditions\cite{clemgonzalez1}
\begin{equation}
\label{eq:dcsmconditions}
|\frac{dH_{\rm n+1}}{dx}|\leq J_{c\perp} \quad ; \quad
|\frac{d\alpha_{\rm n+1}}{dx}|\leq k_{c\parallel}\; .
\end{equation}
Here, $\alpha (x)$ is used for the angle between the local magnetic field vector and a fixed axis.

On making use of the possiblity of expressing the Lagrangian in convenient generalized coordinates, the OC functional may be written
\begin{equation}
\label{eq:hamdcsm}
{\cal C}[\vec{H}_{\rm n+1}(\vec{x})]
=\frac{1}{2}\int_{\Omega}\! H_{\rm n+1}^{2}
-2H_{\rm n}H_{\rm n+1}\cos{(\alpha_{\rm n+1}-\alpha_{\rm n})} \; .
\end {equation}
The modulus-angle representation enables to write the statement of Eq.(\ref{eq:dcsmconditions}) as
\[
\frac{dH_{\rm n+1}}{dx}={J_{\perp}}\equiv
f_{\perp}
\quad ; \quad
\frac{d\alpha_{\rm n+1}}{dx}={k_{\parallel}}\equiv
f_{\parallel}
\]
with $(f_{\parallel},f_{\perp})$ a vector belonging to a rectangular region of size $2k_{c\parallel}\times 2J_{c\perp}$ (see Fig.\ref{fig:regions}). Notice that, in this case, the region $\Delta$ itself is depicted. We also recall that, on using the appropriate dimensionless units $\Delta$ may become a square. Thus, if one defines $\lambda\equiv  1/k_{c\parallel}$ as the length scale and $J_{c\perp}\lambda$ as the magnetic field unit, $\vec{f} $ becomes
\begin{eqnarray}
f_{\perp} &\equiv&
u_{h}\quad ; \quad |u_{h}|\leq 1
\nonumber
\\[1ex]
f_{\parallel} &\equiv&
u_{\alpha}\quad ; \quad |u_{\alpha}|\leq 1  \; .
\nonumber
\end{eqnarray}
For simplicity, this convention will be used below. Then, the associated Hamiltonian is
\begin{equation}
\label{eq:dcsmham}
{\cal H}=p_{h}u_{h}+p_{\alpha}u_{\alpha}-
\frac{1}{2}[H_{\rm n+1}^{2}-2H_{\rm n}H_{\rm n+1}
\cos{(\alpha_{\rm n+1}-\alpha_{\rm n})}]
\; ,
\end{equation}
and the maximality condition is again fulfilled by $\vec{J}^{\; *}\in\partial\Delta$. However, as one can also notice in Fig.\ref{fig:regions}, the rectangular shape determines new features in the {\em critical current density vector} behavior. If the momentum vector $\vec{p}$ has non-vanishing components one gets $\vec{f}^{\; *}=\vec{u}^{\; *}=[{\rm sgn}(p_{\alpha}),{\rm sgn}(p_{h})]$, i.e.: $\vec{f}^{\; *}$ lies in a corner of the rectangle. On the other hand, if either $p_{\alpha}$ or $p_{h}$ vanish, $\vec{f}\cdot\vec{p}$ maximality occurs for any vector leaning somewhere on a given side of the rectangle. Then, $\vec{f}^{\; *}$ must be determined calling on some additional condition. As we have indicated in Fig.\ref{fig:regions}, the previous mathematical properties relate to the physical concept of CT, T and C zones introduced by Clem and P\'erez-Gonz\'alez\cite{clemgonzalez1}. Thus, starting with Eq.(\ref{eq:dcsmham}) one gets
\begin{eqnarray}
\frac{dH_{\rm n+1}}{dx}&=&{\rm sgn}(p_{h})
\label{eqndcsm:a}
\\[1ex]
\frac{d\alpha_{\rm n+1}}{dx}&=&{\rm sgn}(p_{\alpha})
\label{eqndcsm:b}
\\[1ex]
\frac{dp_{h}}{dx}&=&H_{\rm n+1}-H_{\rm n}\cos{(\alpha_{\rm 
n+1}-\alpha_{\rm n})}
\label{eqndcsm:c}
\\[1ex]
\frac{dp_{\alpha}}{dx}&=&H_{\rm n+1}H_{\rm n}
\sin{(\alpha_{\rm n+1}-\alpha_{\rm n})}.
\label{eqndcsm:d}
\end{eqnarray}
The solutions of these equations may be classified in terms of the momentum variable behavior

\begin{itemize}

\item[(i)]CT zone ($p_{\alpha}\neq 0, p_{h}\neq 0$)
\noindent

The field penetration is given by
\begin{eqnarray}
H_{\rm n+1}&=&{\rm sgn}(p_{h})x+H_{\rm n+1}(0)
\nonumber
\\[1ex]
\alpha_{\rm n+1}&=&{\rm sgn}(p_{\alpha})x+\alpha_{\rm n+1}(0) \; .
\nonumber
\end{eqnarray}

\item[(ii)]T zone ($p_{\alpha}= 0, p_{h}\neq 0$)
\noindent

We have
\[
H_{\rm n+1}={\rm sgn}(p_{h})x+H_{\rm n+1}(0)
\]
and $\alpha_{\rm n+1}(x)$ is to be determined by the condition $dp_{\alpha}/dx=0$. Then
\[
H_{\rm n+1}(x)H_{\rm n}(x)\sin{[\alpha_{\rm n+1}(x)-\alpha_{\rm n}(x)]}=0\; .
\]

\item[(iii)]C zone ($p_{\alpha}\neq 0, p_{h}= 0$)
\noindent

We have
\[
\alpha_{\rm n+1}={\rm sgn}(p_{\alpha})x+\alpha_{\rm n+1}(0)
\]
and $H_{\rm n+1}(x)$ is to be determined by the condition $dp_{h}/dx=0$. Then
\[
H_{\rm n+1}(x)=H_{\rm n}(x)\cos{[\alpha_{\rm n+1}(x)-\alpha_{\rm n}(x)]}\; .
\]

\end{itemize}

We notice that the modulus and angle profiles display maximum (critical) slope for the so-called CT zone, corresponding to the fact that both the depinning and cutting thresholds have been exceeded in that region. Thus, flux Cutting as well as flux Transport have occured. Within the so-called T and C zones one gets maximum slope for one of the two components of the field and {\em subcritical} behavior for the other. In physical terms, the highly resistive state has only been triggered for a particular direction in $\vec{J}$-space. 

Finally, we want to mention that the full equivalence of our interpretation to the standard DCSM formulation has been tested.\cite{badia2}  In fact, it has been shown that the latter is nothing but the continuum limit of our time-discretized approach.

\begin{figure}
\centerline{
\includegraphics[height=2.0in]{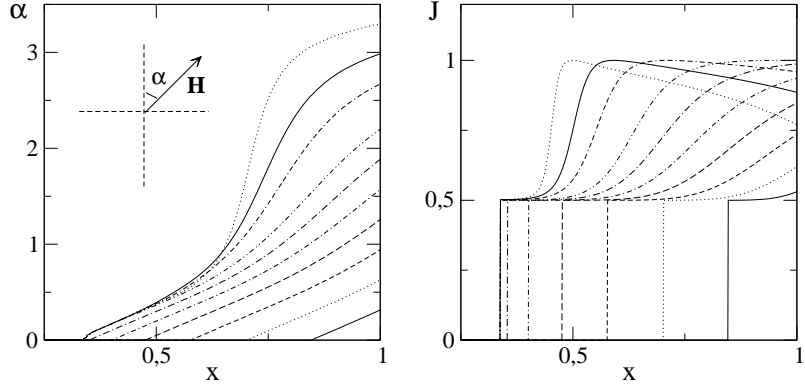}
}
\caption{Rotation angle and current density modulus within the elliptic model. $\alpha$ is given in radians and $J$ in dimensionless units}  
\label{fig:Jellip}
\end{figure}

\subsubsection{Elliptic model}

It is well known that type-II superconductors may display anisotropic physical properties, either related to their crystal structure or artificially induced by a number of processes as thermal or mechanical treatment, irradiation, etc. Thus, we are led to consider the influence of this phenomenon on the critical state. The simplest assumption which one can make is that the modification of the pinning strength along perpendicular directions within the sample can be described by an elliptic region $\Delta$ (see Fig.\ref{fig:regions}). At least, for moderate anisotropy, this may be considered as a first order perturbation of the isotropic model, which provides a good phenomenological theory for many experimental facts.

From the mathematical point of view, the OC equations are also a straightforward modification of the isotropic expressions. Thus, the control system becomes
\begin{eqnarray}
\frac{dH_{\rm n+1,y}}{dx}=f_{y}=\gamma u_y
\label{eq:controlellipy}
\\[1ex]
\frac{dH_{\rm n+1,z}}{dx}=f_{z}=u_z \; ,
\label{eq:controlellipz}
\end{eqnarray}
with $\gamma$ the anisotropy parameter and $\vec{u}$ a vector within the unit disk. This leads to the associated Hamiltonian
\[
{\cal H}=p_{y}\gamma u_{y}+p_{z}u_{z}-\frac{1}{2}
(\vec{H}_{\rm n+1} -\vec{H}_{\rm n})^{2}\; .
\]
In this case, the maximality condition produces new features in the optimum solution. As one can notice in Fig.\ref{fig:regions}, maximum projection does no longer mean $\vec{f}\parallel\vec{p}$ as in the isotropic case. In general, these vectors will be at an angle $0<\beta<\pi /2$, which can be close to $\pi /2$ for $\gamma$ far from unity. Specifically, we get
\[
\vec{f}^{\,*}=\frac{(\gamma^{2} p_{y}^{*},p_{z}^{*})}{\sqrt{\gamma^{2}p_{y}^{*\, 2}+p_{z}^{*\, 2}}}
\; ,
\]
and, thus, the Hamiltonian equations read
\begin{eqnarray}
\frac{dH_{\rm n+1,y}^{*}}{dx}&=&\frac{\gamma^{2}p_{y}^{*}}{\sqrt{\gamma^{2}p_{y}^{*\, 2}+p_{z}^{*\, 2}}}
\label{eqncanslab:a}
\\[1ex]
\frac{dH_{\rm n+1,z}^{*}}{dx}&=&\frac{p_{z}^{*}}{\sqrt{\gamma^{2}p_{y}^{*\, 2}+p_{z}^{*\, 2}}}
\label{eqncanslab:b}
\\[1ex]
\frac{dp_y^{*}}{dx}&=&H_{\rm n+1,y}^{*}-H_{\rm n,y}
\label{eqncanslab:c}
\\[1ex]
\frac{dp_z^{*}}{dx}&=&H_{\rm n+1,z}^{*}-H_{\rm n,z} \; .
\label{eqncanslab:d}
\end{eqnarray}
Notice that this system includes the isotropic model as a particular case ($\gamma=1$) and that excitation fields along the principal axes ($Y$ and $Z$ in this case) produce {\em one-dimensional} critical state models with unequal critical currents as expected.

For illustration of the field penetration properties in this anisotropic model, we include Fig.\ref{fig:Jellip}, in which some aspects of a field rotation experiment are depicted. We have simulated a field consumption process for a field cooled type-II slab with {\em nonmagnetic} initial state ($\vec{H}_{0}=constant\; \hat{z}$) and $\gamma = 0.5$. Notice that, as the magnetic field vector rotates within the sample the current density modulus $J(x)$ varies from 0.5 to 1.

\subsubsection{Pseudo-isotropic model}

\begin{figure}
\centerline{
\includegraphics[height=2.0in]{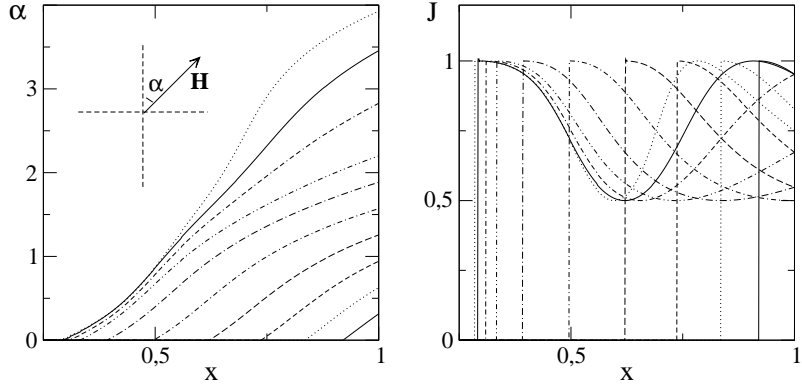}
}
\caption{Rotation angle and current density modulus within the pseudo-isotropic model. $\alpha$ is given in radians and $J$ in dimensionless units}  
\label{fig:Jpseudo}
\end{figure}

In this part, we develop another approximation which is also suited for dealing with anisotropic systems. Within the so-called pseudo-isotropic model, we assume that the restriction region $\Delta$ is a circle, but with variable radius according to the magnetic field orientation with respect to some particular axis within the sample (see Fig.\ref{fig:regions}). This can be written in the form
\[
|\vec{J}| \leq J_{c}\, f(\alpha) \, ,
\]
with $J_c$ a constant and $f(\alpha)$ some function of the angle between the local field and the $Z$ axis. In order to allow comparison with the elliptic model, we introduce the control system
\begin{eqnarray}
\frac{dH_{\rm n+1,y}}{dx}&=&\frac{\gamma H_{\rm n+1,y}^{2}+H_{\rm n+1,z}^{2}}{H_{\rm n+1}^{2}}\, u_y = f_y 
\equiv f(\alpha) u_y
\label{eqnamp2:a}
\\[1ex]
\frac{dH_{\rm n+1,z}}{dx}&=&\frac{\gamma H_{\rm n+1,y}^{2}+H_{\rm n+1,z}^{2}}{H_{\rm n+1}^{2}}\, u_z = f_z 
\equiv f(\alpha) u_z\; .
\label{eqnamp2:b}
\end{eqnarray}
Once more, $\vec{u}$ is a vector within the unit disk. It is apparent that for the particular case of parallel FLL's with magnetic field along principal axes, one gets the same {\em one-dimensional} approximations which arise from Eqs.(\ref{eq:controlellipy}) and (\ref{eq:controlellipz}). For instance, $H_{z}=0\Rightarrow f_{y}=\gamma u_{y}, f_{z}=0$.

The minimization of the cost functional ${\cal C}$ is now attained by maximizing the Hamiltonian
\[
{\cal H}=\frac{\gamma H_{\rm n+1,y}^{2}+H_{\rm n+1,z}^{2}}{H_{\rm n+1}^{2}}\vec{p}\cdot\vec{u}-\frac{1}{2}
(\vec{H}_{\rm n+1} -\vec{H}_{\rm n})^{2}\; .
\]
This leads to the condition $\vec{f}^{\,*}=f(\alpha)\vec{p}^{\,*}/p^{*}$ and one gets the Hamiltonian equations
\begin{eqnarray}
\frac{dH_{\rm n+1,y}^{*}}{dx}&=&\frac{p_{y}^{*}}{p^{*}}\,\frac{\gamma H_{\rm n+1,y}^{*\, 2}+H_{\rm n+1,z}^{*\, 2}}{H_{\rm n+1}^{*\, 2}}
\label{eqncanslab2:a}
\\[1ex]
\frac{dH_{\rm n+1,z}^{*}}{dx}&=&\frac{p_{z}^{*}}{p^{*}}\,\frac{\gamma H_{\rm n+1,y}^{*\, 2}+H_{\rm n+1,z}^{*\, 2}}{H_{\rm n+1}^{*\, 2}}
\label{eqncanslab2:b}
\\[1ex]
\frac{dp_y^{*}}{dx}&=&H_{\rm n+1,y}^{*}-H_{\rm n,y}-2p^{*}
\frac{(\gamma-1) H_{\rm n+1,y}^{*}H_{\rm n+1,z}^{*\, 2}}{H_{\rm n+1}^{*\, 4}}
\label{eqncanslab2:c}
\\[1ex]
\frac{dp_z^{*}}{dx}&=&H_{\rm n+1,z}^{*}-H_{\rm n,z}-2p^{*}
\frac{(1-\gamma) H_{\rm n+1,z}^{*}H_{\rm n+1,y}^{*\, 2}}{H_{\rm n+1}^{*\, 4}}
 \; .
\label{eqncanslab2:d}
\end{eqnarray}

As in the previous model, we illustrate some properties of the magnetic field penetration. Fig.\ref{fig:Jpseudo} depicts the current density variation as the local field rotates within the sample for a {\em nonmagnetic} initial state experiment as in Fig.\ref{fig:Jellip}. Notice that the current density modulus also bounces between the values 0.5 and 1, but with a clearly different structure. Although both models share the same {\em one-dimensional} limits, the actual current distribution $\vec{J}(x)$ is influenced by different factors. Thus, within the elliptic model, the shielding current value is determined by the external field variation. For instance, for the initial rotation steps one gets $J\simeq 0.5$ because the induced current is mainly formed by $J_y$. On the contrary, the magnetic response within the pseudo-isotropic approach is determined by the angle $\alpha$ itself. The actual orientation of $\vec{J}$ is not important. Thus, the initial rotation steps are shielded by $J\simeq 1$ because $\vec{H}$ is still basically directed along $Z$ axis.

\section{COMPARISON TO EXPERIMENT}

In this section we will discuss the application of our theory for the analysis of some relevant experiments  related to the investigation of cross-flow effects. According to the boundary conditions induced by the external drive, they may be classified as: (i) rotation experiments and (ii) crossed-field experiments. In the first case, the applied magnetic field $\vec{H}_{S}$ rotates at some constant and low enough frequency $\omega_{0} $, so as to neglect possible relaxation effects. The field modulus $H_S$ remains constant. In the second case, the orthogonal components of $\vec{H}_{S}$ are cycled for some definite process. For instance, one of them is raised and lowered while the other one is kept constant.

\begin{figure}[tbh]
\centerline{
\includegraphics[height=4.1in]{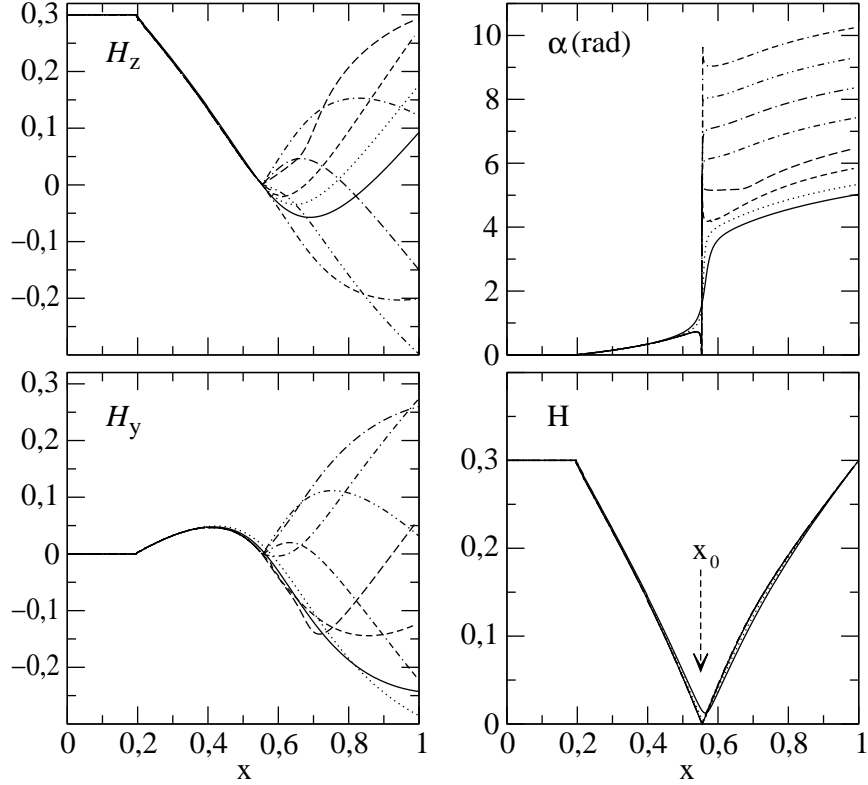}
}
\caption{Simulated field penetration profiles for an isotropic sample under external rotating magnetic field}  
\label{fig:rotiso}
\end{figure}

\subsection{Rotation experiments}

The main features of rotating field experiments\cite{boyer,boyer2,cave} have been well described within the original formulation of the DCSM.\cite{clemgonzalez1} For instance, in the so-called {\em nonmagnetic} initial state, the magnitude of the flux density within the sample is decreased as if vortices were somehow expelled from the sample. Indeed, Clem and Gonz\'alez clarified that such {\em effective} expulsion may be naturally understood as a flux cutting phenomenon between adjacent rotating vortices. Within the DCSM, the following picture was given. The flux density develops a diamagnetic profile toward the sample's midplane, until it becomes zero for some distance $x_0$ (or until the midplane is reached). In the case that $x_{0}>0$ exists, the field modulus profile reaches a stationary V-shape as soon as $x_0$ is defined.

\begin{figure}[tbh]
\centerline{
\includegraphics[height=4.1in]{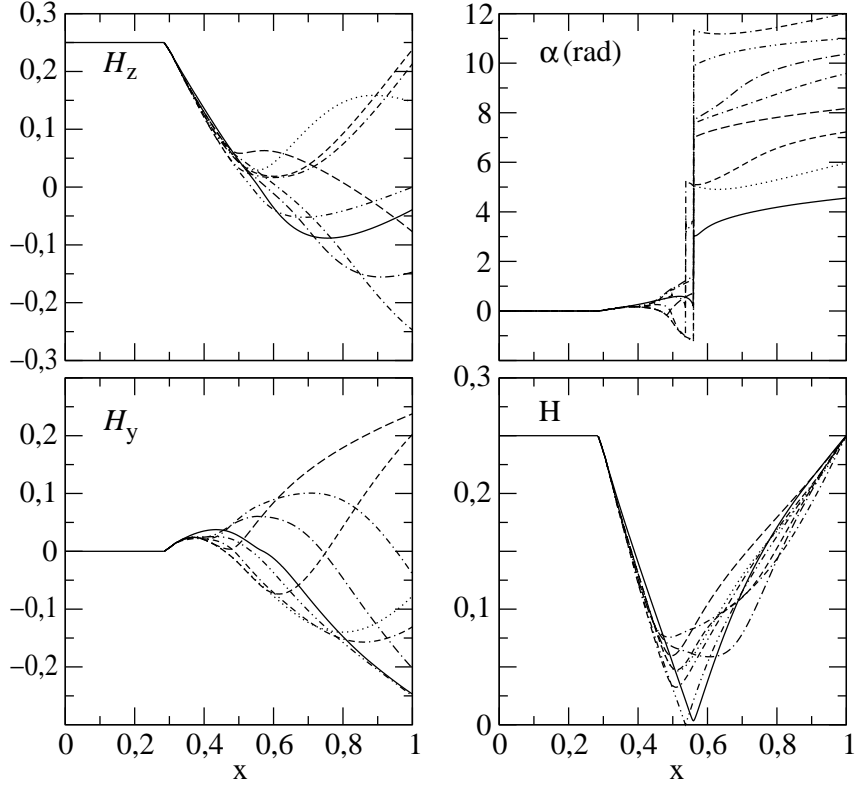}
}
\caption{Simulated field penetration profiles for an anisotropic sample under external rotating magnetic field}  
\label{fig:rotaniso}
\end{figure}

We showed that the behavior described above may also be obtained in the framework of our theory for the isotropic model.\cite{badia1} It was remarked that the decoupling point $x_0$ is sharply defined by the process itself, and that further rotation leaves the field modulus unchanged. This behavior has been emphasized here (see Fig.\ref{fig:rotiso}). We have plotted a number of profiles subsequent to the issue of $x_0$. Notice that $H_{y}(x)$ and $H_{z}(x)$ are frozen for $x<x_0$. Notice also that, though these components evolve for $x>x_0$, one gets a frozen modulus profile and nearly linear penetration  of the rotation angle $\alpha (x)$. Physically, the process may be described as the separation of vortices in two groups, one which rigidly pins to the sample (inner vortices), and another one which frictionally rotates relative to the sample while keeping a constant density profile (outer vortices).  

\begin{figure}[tbh]
\centerline{
\includegraphics[height=3.5in]{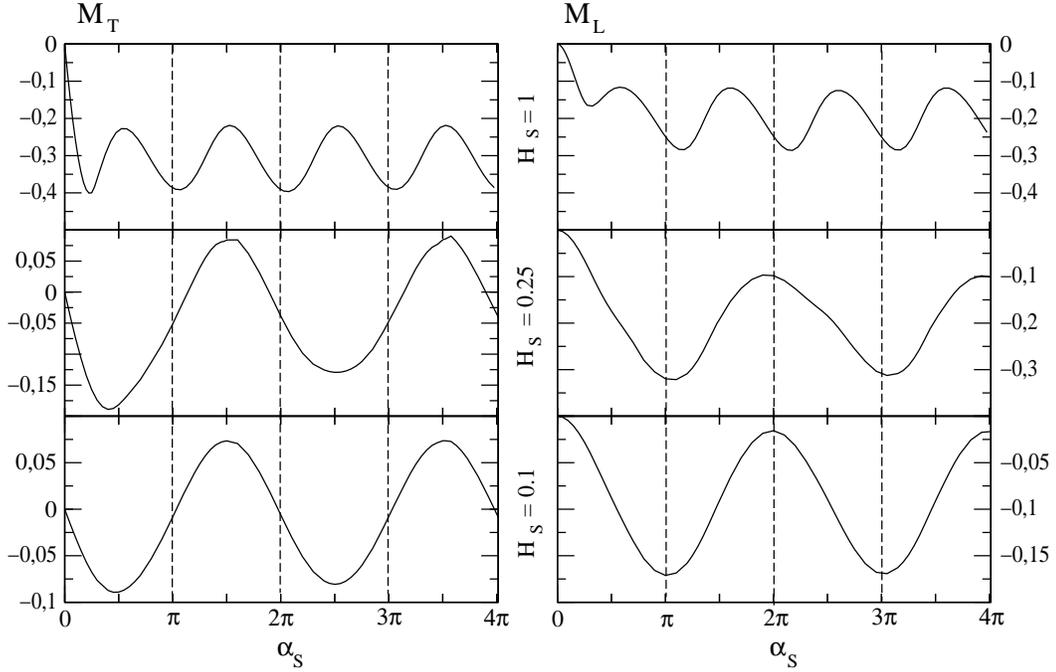}
}
\caption{Simulated magnetization components for an anisotropic sample under external rotating magnetic field}
\label{fig:rotaclem}
\end{figure}

Finally, we will show that, as it was already guessed in Ref.\onlinecite{boyer2}, the appearance of higher frequency oscillations ($2\omega_{0}$) in the magnetic moment components can be ascribed to anisotropy in the current flow. Fig.\ref{fig:rotaniso} depicts our simulations for the field penetration profiles in a rotation experiment as calculated by the pseudo-isotropic model defined in the previous section. We want to notice the appearance of a V-shape modulus profile. However, by contrast to the isotropic case, this structure keeps no longer stationary. Further rotation produces a complex two minima structure. A second turnabout ($2\pi < \omega_{0} < 4\pi$) defines a new V structure and the process goes on in a periodic fashion. As regards the angle penetration profile $\alpha (x)$, a multiple step structure is observed. The definition of decoupling points is accompanied by jumps in the rotation angle. Subsequent steps produce a negative angle for inner points, which is interpreted as a counter rotation of flux lines. The double frequency signal, which was reported in the early 80's by Boyer and co-workers is straightforwardly related to the physical mechanisms described above. In Fig.\ref{fig:rotaclem} we have plotted the averaged magnetization components both longitudinal ($M_L$) and transverse ($M_T$) to the applied field $\vec{H}_S$. The plot collects the evolution $M_{L,T}(\alpha_{S})$ with respect to the applied rotation angle for three different values of the field modulus ($0.1\; , 0.25\; , 1$ in our dimensionless units). Note that, for the lowest field value, $M_{L,T}$ basically display a stationary harmonic behavior at the sample's rotation frequency $\omega_{0}$. As $H_S$ increases, one can readily observe a frequency mixing phenomenon. Eventually, for high enough fields, $\vec{M}$ becomes a rotating vector at frequency $2\omega_{0}$. 

The rotational scenario of field cooled anisotropic samples with nonmagnetic initial state is established as follows. Two groups of vortices exist: a rigid core well within the sample and a region of flux tubes below the surface, which undergo complex transport and cutting phenomena as rotation proceeds. The rigid core contributes as a harmonically oscillating magnetic moment when analyzed in a reference frame at rest with respect to the applied field. On the other hand, the {\em active region} of vortices close to the surface contributes with an essentially double frequency signal. For low enough fields ($H_{S}=0.25$) the rigid core nearly spans over the whole sample and the response is linear. For high enough fields ($H_{S}=1$) the rigid core has disappeared and the response is at $2\omega_{0}$. Intermediate values hold a frequency mixing, corresponding to the superposition of both effects.

\subsection{Crossed field experiments}

For illustration of the characteristic phenomena in the crossed field configuration, we will discuss the results within Refs.\onlinecite{park,fisher}. Hereafter, they will be named {\em Park's} and {\em Fisher's} experiment respectively. In both cases, the superconductor is firstly subjected to a magnetic field along a given direction (f.i.: $H_{zS}$). Subsequently, this component is fixed and the orthogonal one ($H_{yS}$) is cycled while recording the magnetic moment.

\begin{figure}
\centerline{
\includegraphics[height=3.2in]{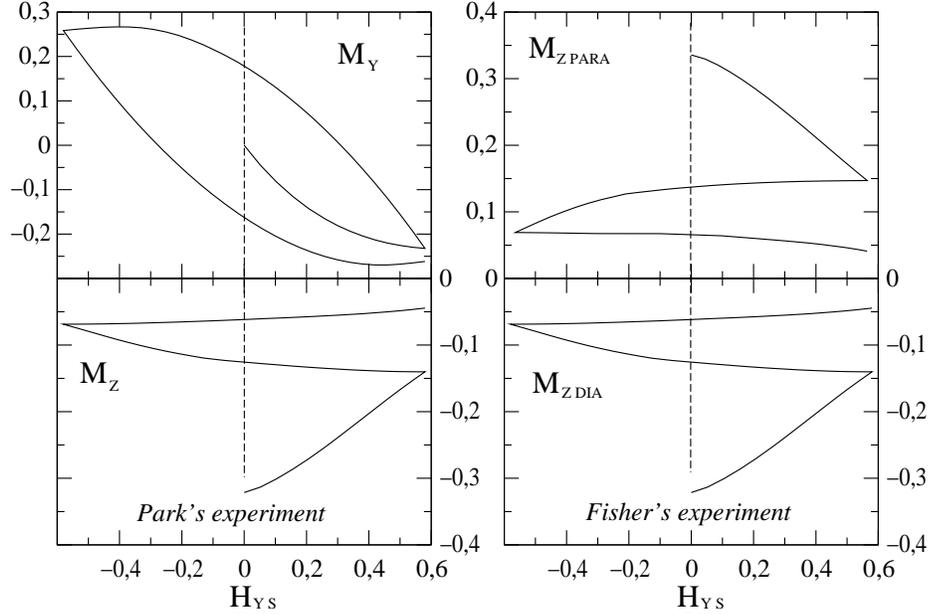}
}
\caption{Simulated magnetization processes for crossed field experiments in the configurations used by Park and Fisher}  
\label{fig:crossed}
\end{figure}

Park and co-workers measured $M_y$ and $M_z$ for high-T$_c$ crystals. Their results show that $M_z$ vanishes as $H_{yS}$ is cycled. Additionally, a more or less conventional loop $M_y(H_{yS})$ is observed. However, the loop fails to close after a field cycle has been completed. Both features are reproduced by the application of our theory within the isotropic model (see Fig.\ref{fig:crossed}).

Finally, the {\em symmetric supression} of the magnetic moment $M_z$ by cycling $H_{yS}$ reported by Fisher may also be explained within the isotropic hypothesis. In Fig.\ref{fig:crossed} we display our simulations of the diminishing behavior of $M_z$, both for the para- and diamagnetic states. In the first case, $H_{zS}$ was cycled up to a positive magnetic moment state, and then fixed. In the second case, $H_{zS}$ was raised up to a definite negative magnetic moment state. 

The underlying physical mechanisms responsible for the described behavior have been extensively discussed in Ref.\onlinecite{badia2}. Basically, owing to the rectriction $|\vec{J}|\leq J_c$, the shielding capability related to $H_z$ is decreased when a considerable amount of current is required for minimizing $\int\! (H_{\rm n+1,y}-H_{\rm n,y})^{2}$, which is the dominant process in the configurations discussed.
\section{CONCLUDING REMARKS}
The fundamental subject of our work is that a number of observables related to experiments with crossing vortices in hard superconductors may be well described in terms of critical state principles. In particular, we have shown that simulated rotation and crossed field experiments both for isotropic and anisotropic systems nicely reproduce the actual measurements.

It is shown that Bean's model for parallel vortices can be thought as a particular case of a general critical state theory. The main ingredients of the theory, which is given in the form of a variational principle, are the magnetic field inertia (Faraday's law) and a very high level of entropy production for moving vortices when the threshold for static configurations is exceeded. The arising constrained minimization problem has been solved by optimal control\cite{oc,leitmann} techniques, which nicely fit the physical statement. However, we want to notice that other methods can be applied to solve the problem posed in Eq.(\ref{eq:varpri}). In particular, we suggets the use of the so-called {\em direct methods}\cite{burden} in which a family of appropriate base functions for the solution generates a problem in matrix form as in standard finite-element theories.

Although many relevant experimental facts can be explained within our theory, several approximations have been used in order to keep the mathematical intricacy to the lowest degree. An improved quantitative description would require the consideration of finite size effects, the influence of equilibrium magnetization, and the finite value of the flux-flow resistivity. These are merely technical aspects which can be incorporated by using OC for partial differential equations, a definite $\vec{B}(\vec{H})$ relation and a dissipation term in the variational quantity.
\section*{ACKNOWLEDGMENTS}
Financial support from Spanish CICYT (projects MAT99-1028 and  BFM-2000-1066/C0301) is acknowledged.
\appendix

\section{A BRIEF SURVEY ON OPTIMAL CONTROL}
\label{a1}

Control theory is a branch  of dynamical systems theory in which the usual ingredients are a family of evolution differential equations of the system containing the time, some phase space coordinates representing the states of the system, and extra control variables modelling some external action applied over it. The applications of the theory have been mainly developed in the field of Engineering. Within this realm, the controls are the representation of some manual or automatized action and, thus, affected by limitations. In mathematical terms, the controls belong to a bounded set.

Self-organized critical systems, which are a flourishing research area in Physics, seem to be a natural candidate for the application of this mathematical tool. In this case, the control variables and limitations are related to physical interactions. In particular, the magnetic properties of hard type-II superconductors rely on the interaction between flux tubes and a maximum force from the underlying pinning structure. 
 
In order to develop the optimal control theory in a brief but self-contained presentation, one should begin by recalling some concepts on the classical variational calculus (see Ref.\onlinecite{leitmann} for a comprehensive and detailed exposition of the topic). Given a Lagrangian function
of position and velocity $L(\vec{x},{d\vec{x}}/{dt})$, the problem of minimizing the action
integral $\int _a^b {L(\vec{x},{d\vec{x}}/{dt})} dt$ for a curve ${\vec{x}}(t)$ with fixed
endpoints $\vec{x}(a)=\vec{x}_a$, $\vec{x}(b) = \vec{x}_b$ generates, by applying a first order variation
$\delta {\vec{x}}(t)$ of the curve and after integration by parts, the well-known
Euler-Lagrange equations
\[
\frac{d}{dt} \left( \frac{\partial L}{\partial v^i} \right) = \frac{\partial
L}{\partial x^i} 
\quad ; \quad 
v^i = \frac{dx^i}{dt} \; .
\]
Above, the boundary conditions which imply the conditions $\delta {\vec{x}}(a)=\delta {\vec{x}}(b)=0$ have been invoked in the integration by parts. For other boundary conditions, these terms do not vanish authomatically, and some
extra {\em transversality} conditions appear. For example, if the final point is not fixed but belongs to some hypersurface $S \subset{\relax{\rm I\kern-.18 em R}}^n$, then its first order variation
$\delta \vec{x}(b)$ is tangent to the hypersurface, and therefore the extra condition is
\[
\frac {\partial L}{\partial {\vec v}} \perp S
\]

 The Hamiltonian version of the classical variational calculus, so important in the process of quantization of the fundamental physical theories, and also in the study of many dynamical systems, will be obtained below using a kind of trick that will be useful later on. Let us consider that coordinates $\vec v$ are independent of coordinates $\vec x$, and afterwards let us impose the constraint $\vec{v}(t) = {d\vec{x}}/{dt}$. In this case, on using the method of Lagrange multipliers for constrained minimization problems, we must use the action integral $\int _a^b {p_0 L(\vec{x},\vec{v}\,) + {\vec p}\cdot({d{\vec x}}/{dt} - {\vec v})} dt$, where $\vec p$ are the Lagrange
multiplier variables, and $p_0$ is a positive constant for minimizing problems. The Euler-Lagrange system for this extended Lagrangian comprises three families of equations, one for each base variable $(\vec{x},\vec{v},\vec{p}\,)$. We get
\[
\frac{dp_i}{dt} = \frac{\partial L}{\partial x^i}
\]
for first order variations in $\vec x$,
\[
0=\frac{dx^i}{dt} - v^i
\]
for first order variations in $\vec p$, and
\[
0=\frac{\partial L}{\partial v^i} - p_i
\]
for first order variation in $\vec v$. 

The last equations define the Legendre transformation between velocity $(\vec{x},\vec{v}\,)$ and phase $(\vec{x},\vec{p}\,)$ spaces. For most cases in the context of classical mechanics, this map is an isomorphism and can be inverted to solve for $\vec{v}$ as a function of $\vec{x}$ and $\vec{p}$. This relates straightforwardly to the condition ${\rm det}(\partial^{2} L/\partial v^i \partial v^j)\neq 0$. Then, it is a simple exercise to check that the other two families of equations can be written in the form
\[
\frac{dp_i}{dt} = -\frac{\partial H}{\partial x^i}
\quad ; \quad 
\frac{dx^i}{dt} =
\frac{\partial H}{\partial p_i}
\]
where $H(\vec{x},\vec{p}\,) =\vec{p}\cdot\vec{v}(\vec{x},\vec{p}\,) - L(\vec{x},\vec{v}(\vec{x},\vec{p}\,))$ is the usual Hamiltonian.

For some {\em pathological} cases, the Lagrangian function is not regular and the Legendre map is not invertible. However, a Hamiltonian theory can still be constructed, but the true Hamiltonian equations live in a mixed velocity-phase space, with independent $(\vec{x},\vec{v},\vec{p}\,)$ coordinates. In such problems, one of the families of equations (those determining $\vec v\,$) is not differential but algebraic. One has a so-called DAE system (Differential Algebraic Equations)
\[
\frac{dp_i}{dt} = -\frac{\partial H}{\partial x^i}
\quad ; \quad
\frac{dx^i}{dt} = \frac{\partial H}{\partial p_i}
\quad ; \quad 
0=\frac{\partial H}{\partial v^i}
\]
with $H(\vec{x},\vec{v},\vec{p}\,) = {\vec{p}}\cdot{\vec{v}} - L(\vec{x},\vec{v}\,)$.

Let us now introduce a {\em control system} for the velocities
\[
\frac{d\vec{x}}{dt} = \vec{f}(\vec{x},\vec{u}\,) \; .
\]
Here, the coordinates $\vec{u}$ are the so-called controls, in general, taking values within some subset $\Delta\subset{\relax{\rm I\kern-.18 em R}}^k$. These extra coordinates may be modelling some external action on the system or also some intrinsic limitation as it is the case of the critical current density in superconductors. Let us also assume that we want to minimize a functional $\int L(\vec{x},\vec{u}\,) dt$. We notice that the control system may be thought as a parametric description of some subset of the velocity space, and therefore as a constraint for the minimization problem. This constraint directly relates to the singular behavior of the matrix $\partial^{2} L/\partial v^i \partial v^j$ and the method for pathological systems must be applied. One must consider the Hamiltonian equations for $H(\vec{x},\vec{u},\vec{p}\,) = {\vec p}\cdot{\vec f}(\vec{x},\vec{u}\,)-L(\vec{x},\vec{u}\,)$, i.e.
\begin{eqnarray}
\frac{dx^i}{dt} = \frac{\partial H}{\partial p_i}=f^i(\vec{x},\vec{u}\,)
\nonumber
\\
\frac{dp_i}{dt} = -\frac{\partial H}{\partial x^i}= \frac{\partial L}{\partial x^i}-
{\vec p}\cdot\frac{\partial {\vec f}}{\partial x^i}
\nonumber
\\
0=\frac{\partial H}{\partial u^a}= {\vec p}\cdot\frac{\partial {\vec f}}{\partial u^a}-
\frac{\partial L}{\partial u^a} \; .
\nonumber
\end{eqnarray}
Note that the last equation is a first order necessary condition of
minimality relating only to stationary internal points of the region $\Delta$. For general problems with bounded control spaces, a stronger condition for determining the solution $\vec{u}^{\,*}$ is required. This is given by the so-called {\em maximum principle of Pontryagin}\cite{oc,leitmann}, which we state without proof here
\[
\vec{u}^{\,*} \ni H(\vec{x},\vec{u}^{\,*},\vec{p}) =  
\max_{\vec{u}\in\Delta}{H(\vec{x},\vec{u},\vec{p}\,)} \; .
\]
As far as this condition can be explicitly solved, we get $\vec{u}^{\,*}(\vec{x},\vec{p}\,)$, and after back substitution into the other equations we get a system of first order ordinary differential equations. As regards the boundary conditions, they can be simple as in the case of fixed endpoints for $\vec x$, or they can include transversality conditions, as pointed before. For example, when the final set for ${\vec x}$ is an hypersurface of ${\relax{\rm I\kern-.18 em R}}^n$, the transversality condition means that $\vec p$ is perpendicular to
the hypersurface (in particular, free final point ${\vec x}$ means that the hypersurface is ${\relax{\rm I\kern-.18 em R}}^n$ itself and implies a vanishing final momentum).
Sometimes the problem does not have a defined final point $b$; then, the associated transversality condition is the vanishing of the Hamiltonian at this unknown point.\cite{leitmann}

Just for simplicity, the previous exposition has been done within the scenario of particle dynamics. Of course, it can be extended to continuous systems by using the fields as dynamical variables and defining the corresponding Lagrangian and Hamiltonian densities.

\vspace{2.5cm}

{{\hspace{-.5cm}\fboxsep
1.ex\fbox{\parbox{0.95\textwidth}{ 
{\em This work is dedicated to Professor D. Gonz\'alez, who has been devoted to Low Temperature Physics, on the occasion of his retirement.
}
}}}}


\begin{thebibliography}{9}

\bibitem{bean} {C. P. Bean}, {\em Phys. Rev. Lett.} {\bf 8}, 250 (1962); 
{\em Rev. Mod. Phys.} {\bf 36}, 31 (1964).

\bibitem{richardson} {R. A. Richardson, O. Pla, and F. Nori}, {\em Phys. Rev. Lett.} {\bf 72}, 1268 (1994).

\bibitem{campbellevetts} {A. M. Campbell and J. E. Evetts}, {\em Critical currents in superconductors} (Taylor \& Francis, London 1972)

\bibitem{boyer} {R. Boyer and M. A. R. LeBlanc}, {\em Sol. Stat. Comm.} {\bf 24}, 261 (1977).
\bibitem{boyer2} {R. Boyer, G. Fillion, and M. A. R. LeBlanc}, {\em J. Appl. Phys.} {\bf 51}, 1692 (1980).
%
\bibitem{cave} {J. R. Cave and M. A. R. LeBlanc}, {\em J. Appl. Phys.} 
{\bf 53}, 1631 (1982).
%
\bibitem{clemgonzalez1} {J. R. Clem and A. P\'erez-Gonz\'alez}, {\em Phys. 
Rev. B} {\bf 30}, 5041 (1984).
%
\bibitem{voloshin} {I. F. Voloshin, A. V. Kalinov, S. E. Savel'ev, L. 
M. Fisher, V. A. Yampolski\u{\i}, and F. P\'erez-Rodr\'{\i}guez}, 
{\em JETP} {\bf 84}, 592 (1997).
%
\bibitem{bosspri} {A. Bossavit}, {\em IEEE Trans. on Magnetics} {\bf
30}, 3363 (1994); L. Prigozhin, {\em J. Comput. Phys.} {\bf 129}, 190 (1996).
%
\bibitem{bhagwat} {K. V. Bhagwat, S. V. Nair, and P. Chaddah}, {\em Physica C} {\bf 227}, 176 (1994).
%
\bibitem{krasnov} {V. M. Krasnov and V. V. Ryazanov}, {\em Physica C} {\bf 297}, 153 (1998).
%
\bibitem{badia1} {A. Bad\'{\i}a and C. L\'opez}, {\em Phys. Rev. Lett.} {\bf 87}, 127004 (2001).
%
\bibitem{badia2} {A. Bad\'{\i}a and C. L\'opez}, {\em Phys. Rev. B} {\bf 65}, 104514 (2002).
%
\bibitem{prigogine} {I. Prigogine}, {\em Introduction to 
Thermodynamics of Irreversible Processes} (North-Holland, Amsterdam, 
1967).
%
\bibitem{landau} {L. D. Landau and E. M. Lifshitz}, {\em Theoretical Physics}, Vol. Electrodynamics of Continuous Media, 2nd ed. (Pergamon, Oxford 1984).
%
\bibitem{oc} {L. S. Pontryagin, V. Boltyanski\u{\i}, R. Gramkrelidze, 
and E. Mischenko}, {\em The Mathematical Theory of Optimal Processes} 
(Wiley Interscience, New York, 1962).
%
\bibitem{baltaga} {I. V. Baltaga, N. M. Makarov, V. A. Yampol'skii, L. M. Fisher, N. V. Il'in, and I. F. Voloshin}, {\em Phys. Lett. A} {\bf 148}, 213 (1990).
%
\bibitem{park} {S. J. Park, J. S. Kouvel, H. B. Radousky, and J. Z. 
Liu}, {\em Phys. Rev. B} {\bf 48}, 13998 (1993).
%
\bibitem{fisher} {L. M. Fisher, K. V. Il'enko, A. V. Kalinov, M. A. R. LeBlanc, F. P\'erez-Rodr\'{\i}guez, S. E. Savel'ev, I. F. Voloshin, and V. A. Yampolski\u{\i}}, {\em Phys. Rev. B} {\bf 61}, 15382 (2000).
%
\bibitem{leitmann} {G. Leitmann}, {\em The Calculus of Variations and 
Optimal Control}, Mathematical Concepts and Methods in Science and 
Engineering, Vol. 24, edited by A. Miele (Plenum Press, New York, 
1981).
%
\bibitem{burden} {R. L. Burden and J. D. Faires}, {\em Numerical Analysis}, 7th ed. (Brooks/Cole Publishing Co. 2001).
%
\end{thebibliography}
\end{document}